\documentclass[aps,prb,reprint,superscriptaddress,amsmath]{revtex4-1}

\usepackage{graphicx}

\usepackage[normalem]{ulem}

\newcommand{\bo}[1] 
{
	\textbf{#1}
}

\newcommand{\ket}[1]
{
	\vert {#1} \rangle
}

\newcommand{\updown}
{
  \uparrow \downarrow
}

\newcommand{\downup}
{
  \downarrow \uparrow
}

\begin{document}

\title{Positron surface state as a new spectroscopic probe for
characterizing surfaces of topological insulator materials}

\author{Vincent Callewaert} \email{vincent.callewaert@uantwerpen.be}
\affiliation{Department of Physics, Universiteit Antwerpen, Antwerpen 2020,
Belgium}

\author{K. Shastry} \affiliation{Department of Physics, University of Texas
at Arlington, Arlington, Texas 76019, USA}

\author{Rolando Saniz} \affiliation{Department of Physics, Universiteit
Antwerpen, Antwerpen 2020, Belgium}

\author{Ilja Makkonen} \affiliation{Department of Applied Physics, Aalto
University School of Science, P.O.\ Box 15100, FI-00076 Aalto, Espoo, Finland}

\author{Bernardo Barbiellini} \affiliation{Department of Physics, Northeastern
University, Boston, Massachusetts 02115, USA}

\author{Badih A.\ Assaf} \affiliation{Department of Physics, Northeastern
University, Boston, Massachusetts 02115, USA}
\affiliation{D\'epartement de physique, Ecole Normale Sup\'erieure, CNRS, PSL Research
University, 24 rue Lhomond, 75005 Paris, France}

\author{Donald Heiman} \affiliation{Department of Physics, Northeastern
University, Boston, Massachusetts 02115, USA}

\author{Jagadeesh S.\ Moodera} \affiliation{Department of Physics, MIT,
Cambridge, Massachusetts 02139, USA}
\affiliation{Francis Bitter Magnet Laboratory, MIT, Cambridge, Massachusetts,
02139, USA}

\author{Bart Partoens} \affiliation{Department of Physics, Universiteit
Antwerpen, Antwerpen 2020, Belgium}

\author{Arun Bansil} \affiliation{Department of Physics, Northeastern
University, Boston, Massachusetts 02115, USA}

\author{A.\ H.\ Weiss} \affiliation{Department of Physics, University of Texas at
Arlington, Arlington, Texas 76019, USA}

\begin{abstract}
Topological insulators are attracting considerable interest due to
their potential for technological applications and as platforms for exploring
wide-ranging fundamental science questions. In order to exploit, fine-tune,
control and manipulate the topological surface states, spectroscopic tools which
can effectively probe their properties are of key importance. Here, we
demonstrate that positrons provide a sensitive probe for topological states, and
that the associated annihilation spectrum provides a new technique for
characterizing these states.  Firm experimental evidence for the existence of a
positron surface state near Bi$_2$Te$_2$Se with a binding energy of $E_b = 2.7
\pm 0.2 \, \text{eV}$ is presented, and is confirmed by first-principles
calculations. Additionally, the simulations predict a significant signal
originating from annihilation with the topological surface states and shows the
feasibility to detect their spin-texture through the use of spin-polarized
positron beams. 
\end{abstract}

\date{\today}


\maketitle

\section{Introduction}
Quickly after their initial discovery, topological insulators (TIs) were
recognized to hold significant potential for new technological applications and
as playground for fundamental physics~\cite{Hasan2010, *Bansil2016, *Moore2010,
*Mellnik2014, *Mourik2012}. An intrinsic challenge with TIs, which arises due to
the fact that their interesting properties originate from Dirac states located
in a nanoscopic layer near the surface, remains to separate the fingerprint of
the topological surface states from the bulk behaviour of the sample. Highly
surface sensitive techniques such as angle resolved photoemission spectroscopy
and scanning tunnelling microscopy have thus proven to be an indispensable tool
to establish the existence of the gapless states in several systems and to
confirm various of the predicted quasi-particle properties~\cite{Hsieh2008,
*Hsieh2009, *Roushan2009, *Xia2009}.\\
In this article, we demonstrate that positrons provide a highly
surface sensitive probe for the topological Dirac states. Since positron
annihilation spectroscopy (PAS) techniques, with measurements of the 2D angular
correlation of the annihilation radiation (2D-ACAR) in particular, are well
suited to measure both the low and high momentum components of the annihilating
electronic states without complication of matrix element effects, they can
provide useful information on the Dirac state orbitals.  Our calculations show
that spin-polarized positron beams can additionally resolve the spin-textures
associated with the topological states, owing to the predominant annihilation
between particles with opposite spins~\cite{Berko1964}. \\
In section~\ref{section:Experiments}, we present the experimental evidence for
the existence of a bound positron state at the surface of the TI Bi$_2$Te$_2$Se
and the measured binding energy~\cite{Shastry2015}. Section~\ref{section:Theory}
contains a discussion of the theory and computational details  used in our first
principles investigation. In section~\ref{section:ResultsTheory}, we show that
the theory confirms the experimental interpretation and predicts a significant
overlap between the positron and the topological states. We also demonstrate
that spin-polarized positron measurements can reveal the spin-structure at
the surface. In section~\ref{section:Conclusion} we summarize the results and
discuss possible applications and advantages of PAS over other spectroscopic
techniques.

\section{\label{section:Experiments} Experimental results}
\begin{figure}
\includegraphics[width=86mm]{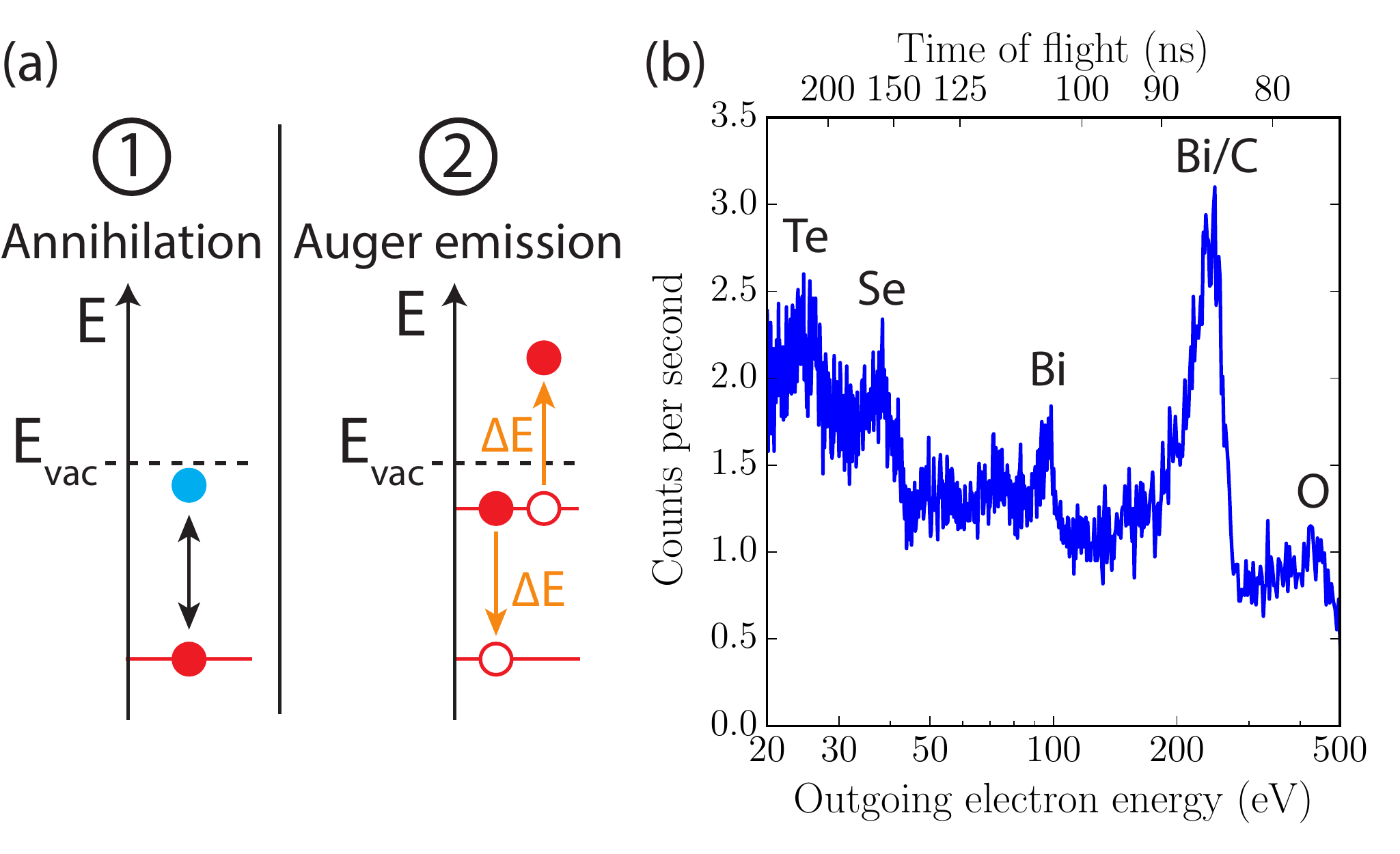}
\caption{\label{fig:PAESExperiment}  (a) Schematic representation
of the PAES mechanism. In the first step, a positron (blue) annihilates with an
electron (red) occupying a core level and creates a highly unstable hole. In the
second step, an electron from a higher level fills this hole and transfers the
energy difference between the two levels to a second electron. If the energy
difference is sufficiently large, and the second electron is close enough to the
surface, it can traverse the surface dipole and escape from the sample. The
measured outgoing electron energy corresponds with the transferred energy in the
Auger process minus the energy difference between the second electron's state
and the vacuum level. (b) Results of the PAES measurements on the Bi$_2$Te$_2$Se
sample in which Auger signals from the different elements are indicated.}
\end{figure}

\begin{figure}
\includegraphics[width=86mm]{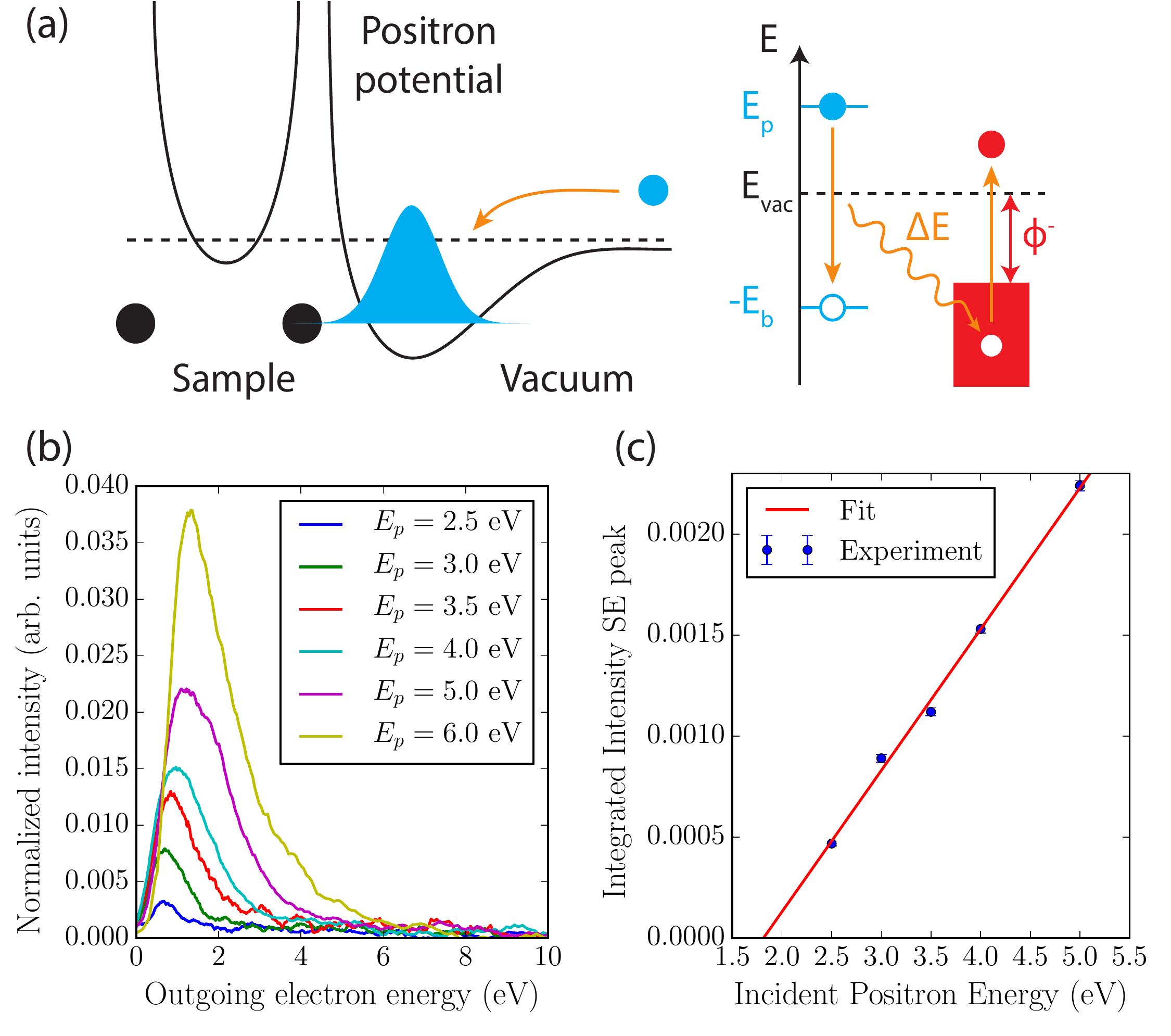}
\caption{\label{fig:AMPSExperiment}  (a) Schematic representation
of the AMPS mechanism. The left part of the diagram shows the incident positron
(blue) that drops in the image potential well. In this process, the positron
transfers an energy $\Delta E$, determined by the incident kinetic energy $E_p$
and the binding energy of the SS $E_b$, to an electron of the system through a
virtual photon, as indicated in the right part of the figure. If the energy
difference is larger than the electronic workfunction $\phi^-$, the electron can
escape to the vacuum. (b) The measured low-energy Auger signals for the
Bi$_2$Te$_2$Se sample. The outgoing electron energy is determined by the
transferred energy $\Delta E$ minus the required energy to escape from the
sample. The different lines show the result for varying energies of the incident
positrons. (c) The integrated peak amplitudes of the low-energy Auger signal
associated with the AMPS mechanism as a function of the incident positron
energy.}
\end{figure}

Our Bi$_2$Te$_2$Se films are grown by molecular beam epitaxy on Si (111). The
substrates are etched in hydrofluoric acid prior to loading in vacuum. A
stoichiometric 2:2:1 Bi:Te:Se flux ratio is used. The substrate temperature is
fixed at 200 $^\circ$C during the growth. The films used in this study are
typically 40 nm thick. A 100 nm Se cap is then deposited, in-situ, on the sample
surface after cooling down the substrate to room temperature. The capping layer
protects the film surface from oxidation and atmospheric contaminants. \\ 
X-ray diffraction (XRD) is systematically used to characterize the
samples, as briefly discussed in ref.~\onlinecite{Assaf2013}. The c-axis lattice
constant for the film used in this work is found to be equal to $30.10 \pm
0.03~\text{\AA}$. Energy dispersive X-ray spectroscopy confirmed stoichiometry
within a 5\% error on samples resulting from an identical growth.\\ 
The samples are then transferred to the experimental positron chamber. In order
to decap the samples, the protective Se layer is evaporated under UHV
conditions, prior to the positron annihilation experiment. A heater button is
placed behind the sample in a holder and a suitable current was passed to heat
the sample for 20 minutes at 200 $^\circ$C. This procedure is similar to the
decapping sequence used in ref.~\onlinecite{Zhang2012}. The technical details
concerning the setup of the positron experiments can be found in
ref.~\onlinecite{Shastry2015} and references therein.\\ 
Positrons annihilate predominantly with the valence electrons but the small
fraction that annihilates with core electrons produces highly unstable core
holes, which are filled by the Auger process. Therefore, if positrons annihilate
in a surface state (SS), positron-induced Auger-electron spectroscopy (PAES)
provides a particularly clean method to determine the composition of the
surface, free from a secondary electron background~\cite{Weiss1988}. A schematic
picture of the process is drawn in figure~\ref{fig:PAESExperiment}(a). Results
of PAES experiments from the TI Bi$_2$Te$_2$Se surface are shown in
figure~\ref{fig:PAESExperiment}(b), where signals from Bi, Te, Se, C and O can
be identified; the latter two are caused by the presence of a small
concentration of contaminants adsorbed on the surface~\cite{Shastry2015}. These
results reveal the presence of a bound positron SS. Were this not the case,
positrons would either get trapped between the blocks of quintuple layers (QL)
of the material or would be re-emitted before they annihilate. Since the extent
of one QL block is about $10 \, \text{\AA}$, which corresponds roughly to the
mean free path of a $60 \, \text{eV}$ electron, any Auger signal coming from
below the first QL is too weak to be detected.  Thus, the fact that the
annihilation induced Auger peak intensities are observable is clear evidence
that the positron is in a state localized at the surface at the time it
annihilates.\\
Auger Mediated Positron Sticking (AMPS) experiments provide an independent proof
for the existence of the SS and allow us to determine its binding
energy~\cite{Mukherjee2010}. In the AMPS mechanism, the excess energy from a
positron dropping into the image potential well is transferred to a valence
electron. This can result in the emission of an Auger electron if the energy
difference between the positron SS and the initial state, determined by the
incident positron's kinetic energy, is larger than the electron
workfunction~\cite{Mukherjee2010}. The maximum kinetic energy of the Auger
electrons is then given by $E_{max} = E_p + E_b - \phi^-$, where $E_p$ is the
energy of the incident positron, $E_b$ is the binding energy of the positron
surface state, and $\phi^-$ is the electron workfunction.
Figure~\ref{fig:AMPSExperiment}(a) illustrates the AMPS mechanism schematically.
The observed increase in amplitude of the Auger signal at low energies as the
energy of the incident positrons is increased, is shown in
figure~\ref{fig:AMPSExperiment}(b), and it confirms the presence of the SS.
Knowing the electron workfunction, the binding energy of the SS can be
determined from the positron energy threshold value for Auger electron emission:
$E_{Th} = E_p$ for which $E_{max} = 0$. The linear fit shown in figure
\ref{fig:AMPSExperiment}(c), yields $E_{Th} = 1.8 \, \text{eV}$. Next, by
considering the measured activation energy $E_a = 0.4 \, \text{eV}$ for
positronium (Ps) desorption from the surface~\cite{Shastry2015}, one can
eliminate the electron workfunction using the expression~\cite{Chu1981}, $E_a =
E_b + \phi^- - 6.80 \, \text{eV}$, which gives a binding energy of $E_b = 2.7
\pm 0.2 \, \text{eV}$ (ref.~\onlinecite{Shastry2015}).

\section{\label{section:Theory} Theory and computational details}
Our first-principles calculations are carried out in the zero-positron-density
limit of the two-component electron-positron density functional theory
(2CDFT)~\cite{Chakraborty1983, Boronski1986}. In this limit, which is exact in
the case of a delocalized positron in a perfect crystal or at a surface, the
electron density remains unperturbed by the presence of the positron. The
computations thus consists of an electronic and positronic groundstate
calculation which are performed subsequently.

\subsection{Electronic structure}
The electronic ground state is obtained using the projector-augmented wave (PAW)
method~\cite{Blochl1994} as implemented in the VASP software
package~\cite{Kresse1996A, Kresse1996B, Kresse1999}. Electron
exchange-correlation effects are treated using the Perdew-Burke-Ernzerhof (PBE)
functional~\cite{Perdew1996}, and spin-orbit coupling is included in the
computations. The kinetic energy cutoff for the plane-wave expansion of the
wavefunctions is set to $275 \, \text{eV}$. For the bulk calculations, we use
the rhombohedral unit cell with a $\Gamma$-centered $11 \times 11 \times 11$
$\textbf{k}$-grid in combination with a Gaussian smearing of width $0.1 \,
\text{eV}$. In the surface calculations, we use a slab geometry with a vacuum of
$15 \, \text{\AA}$ to avoid spurious interactions between periodic images. Here,
the calculations are performed with a $\Gamma$-centered $\textbf{k}$-grid with
$11 \times 11 \times 1$ points in the hexagonal unit cell in combination with a
Gaussian smearing of $0.1 \, \text{eV}$. We used the experimental lattice
parameters in all our calculations~\footnote{The distance between the QL blocks
is severely overestimated when using the PBE functional. As positrons are
strongly repelled by the ions, the separation between the QL strongly influences
the value of the positron workfunction and in order to obtain reliable results,
we deem it appropriate to work with the experimental lattice parameters instead.
The lattice parameters only slightly affect the electronic structure as the
results of our bandstructure calculations agree very well with the previously
reported first-principles results~\cite{Dai2012,Wang2011,Chang2011,Lin2011} and
those of ARPES measurements~\cite{Neupane2012,Arakane2012}}.\\

\subsection{Positron state}
The effective potential for the positron in the zero-density limit of the 2CDFT
is determined by the Coulomb interaction with the nuclei, the Hartree
interaction with the electron density and the electron-positron correlation
potential. The latter is usually described with local density 
(LDA) or generalized gradient approximations (GGA), which give
reliable results for bulk systems. A fundamental limitation of these semi-local
approximations is that they always describe the formation of Ps$^-$ in the
limit of a dilute electron gas. In the case of a surface, however, the correct
limit is given by the image potential~\cite{Lang1973} $-1/4(z-z_0)$, where $z$
denotes the distance to the surface and $z_0$ represents the image potential
reference plane. We impose this limit in the vacuum region by considering the
corrugated mirror model~\cite{Nieminen1983}, in which the image potential is
constructed to follow the same isosurfaces as the electron density. In the
vacuum region $z > z_0$, we take the least negative of the LDA
potential~\footnote{We are updating the standard corrugated mirror model for the
potential at the surface \cite{Nieminen1983, Fazleev1995, Fazleev2004} where GGA
corrections \cite{Barbiellini2015} are traditionally not included.} and the
image potential.  The strength of the image potential is given
by~\cite{Nieminen1983}:
\begin{equation}
V_\text{im}(\textbf{r}) = - \frac{1}{4(z_\text{eff}(n^-(\textbf{r})) - z_0)},
\end{equation}
where $n^-(\textbf{r})$ is the electron density and the effective distance to
the surface is determined by:
\begin{equation}
z_\text{eff}(n^-(\textbf{r})) = \int \limits_{z_0}^\infty dz' \, z' \,
  \delta(n^-(\textbf{r}) - \langle n^- \rangle (z')).
\end{equation}
Here, $\langle n^- \rangle$ is the electron density averaged over the planes
parallel to the surface and $\delta$ denotes the Dirac delta function. We
approximate the image potential reference plane $z_0$ by the background edge
position, which is determined by the position outside the surface where the
electron density starts decaying exponentially.\\
We used the MIKA/doppler package~\cite{Makkonen2006} to obtain the positron
ground state. These calculations are performed in an all-electron way in the
sense that a superposition of free atomic core quantities, e.g.\ density and
Hartree potential, are added to the self-consistent valence electron properties.
The Kohn-Sham equations for the positron are solved on a real space grid using a
Rayleigh multigrid implementation~\cite{Torsti2003, Torsti2006}.

\subsection{Electron-positron momentum density}
The goal of the present paper is to investigate whether PAS can be used to
measure the properties of the TI's Dirac states. We thus need to calculate the
electron-positron momentum density, which contains information about a sample's
electronic structure, and determine if it contains a clear fingerprint of the
topological states.\\
 To the best of our knowledge, electron-positron momentum density calculations
in which the electronic wavefunctions are not collinear, have not been studied
in literature. Hence, we present in some detail a generalization of the theory
to the non-collinear case.\\
Spin-polarized positron annihilation measurements exploit the fact that the two
gamma annihilation only occurs for electron-positron pairs in a singlet state.
If one specifies the initial spin of the positron, this translates to saying
that the positron will only annihilate with electrons of the opposite spin. The
magnetization of the electron-positron momentum density along a specified axis
can thus be obtained by taking the difference between spectra obtained by
aligning the positrons parallel and anti-parallel to that axis. As long as the
electron and positron spins are good quantum numbers, i.e.\ they are position
independent, the effect of the spin is easily taken into account by realizing
that the positron will be in a singlet state with exactly half of the electron
states with the opposite spin. In systems where the spin cannot be considered a
good quantum number, however, a more careful examination is required. In
general, we can write the momentum density of the annihilating electron-positron
pairs as~\cite{Makkonen2005, Zubiaga2016}:
\begin{equation} \label{eq:md_general} \begin{split}
  \rho(\textbf{p}) & = 4 \pi r_e^2 c \sum_j g_j \sum_{s_e, s_p} \Bigg \vert \int
  d\textbf{r} \, e^{ - i \textbf{p} \cdot \textbf{r}} \, \hat{S}^s 
  \alpha_{j}(\textbf{r}, s_e; \textbf{r}, s_p) \Bigg \vert^2 \\
\end{split} \end{equation}
where $\vert \alpha_{j} \rangle$ are the natural geminals which diagonalize the
reduced two-body density matrix, sometimes also referred to as electron-positron
pairing wavefunctions, and the $g_{j}$ are their occupation numbers. The spin of
the electron and positron in the geminal are denoted by $s_e$ and $s_p$,
respectively, and $j$ represents a set of quantum numbers (excluding the spin of
the particles). The factor $4 \pi
r_e^2 c$, with $r_e$ the classical electron radius and $c$ the speed of light,
is the annihilation rate constant~\cite{Ferrell1956}. The operator
$\hat{S}^s = 1 - \frac{1}{2} \hat{S}^2$, where $\hat{S}$ is the total spin
operator for the electron-positron pair, projects on the singlet state. For the
purpose of notation as well as practical calculations, it is convenient to
define:
\begin{equation}
  A_{j, s_e, s_p}(\textbf{p}) = \int d \textbf{r} e^{-i \textbf{p} \cdot
  \textbf{r}} \alpha_{j}(\textbf{r}, s_e; \textbf{r}, s_p)
\end{equation} 
as well as the matrix:
\begin{equation}
\Gamma_j(\textbf{p}) = 
  \begin{pmatrix}
    \vert A_{j, \uparrow \downarrow}(\textbf{p}) \vert^2 &
    A_{j, \uparrow \downarrow}(\textbf{p}) 
      A^*_{j, \downarrow \uparrow}(\textbf{p}) \\
    A_{j, \downarrow \uparrow}(\textbf{p}) 
      A^*_{j, \uparrow \downarrow}(\textbf{p}) &
    \vert A_{j, \downarrow \uparrow}(\textbf{p}) \vert^2 
  \end{pmatrix}.
\end{equation}
In measurements with unpolarized positron beams, the positron has statistically
a 50\% chance to be either in the spin-up or spin-down state. In this
case, upon evaluation of eq.~\eqref{eq:md_general}, the off-diagonal terms of
$\Gamma_j(\textbf{p})$ drop since the geminals with opposite spin orientations,
e.g.\ $\alpha_j(\textbf{r}, \uparrow; \textbf{r}, \downarrow$) and
$\alpha_j(\textbf{r}, \downarrow; \textbf{r}, \uparrow)$, are not simultaneously
occupied. The result for the momentum density then becomes: \begin{equation}
\label{eq:md_total}
  \rho(\textbf{p}) = \pi r_e^2 c \sum_j g_j \text{Tr}[\Gamma_j(\textbf{p})],
\end{equation}
where $\text{Tr}[\dots]$ denotes taking the trace. In case the positron beam is
perfectly polarized parallel or anti-parallel to the $z$-axis, we obtain:
\begin{equation}
\rho_z^\uparrow(\textbf{p}) = 2 \pi r_e^2 c \sum_j g_j \vert
A_{j, \uparrow \downarrow}(\textbf{p}) \vert^2, 
\end{equation}
and:
\begin{equation}
\rho_z^\downarrow(\textbf{p}) = 2 \pi r_e^2 c \sum_j g_j \vert
A_{j, \downarrow \uparrow}(\textbf{p}) \vert^2, 
\end{equation}
respectively. The magnetization along the $z$-axis is obtained by taking the
difference between these two spectra, and can conveniently be written as:
\begin{equation} \begin{split}
  \rho_z(\textbf{p}) & = 2 \pi r_e^2 c \sum_j g_j 
  \text{Tr}[\sigma_z \Gamma_j(\textbf{p})],
\end{split} \end{equation}
where $\sigma_z$ denotes the Pauli matrix.  Analogous observations can be made
for a positron polarized along the different axes, thus we can write in general:
\begin{equation} \label{eq:md_magnetization}
  \rho_i(\textbf{p}) = 2 \pi r_e^2 c \sum_j g_j
  \text{Tr}[\sigma_i \Gamma_j(\textbf{p})],
\end{equation}
where $i = \{ x, y, z \}$ and the $\sigma_i$ are the Pauli matrices. A detailed
derivation of the above formulas can be found in the appendix.\\ 
In electron-positron momentum density calculations based on the 2CDFT, one
assumes that the natural geminals can be written in terms of a product of the
electron and positron single particle Kohn-Sham orbitals $\psi^-_{j, s_e}$ and
$\psi^+_{s_p}$, where the positron is assumed to reside in its groundstate, and
the occupation numbers of the electronic orbitals replace those of the natural
geminals $g_j$.  Electron-positron correlation effects are included by
introducing a multiplicative term $\gamma$, i.e.\ the enhancement factor, which
can be state and/or space dependent. We thus have:
\begin{equation}
  \alpha_{j}(\textbf{r}, s_e; \textbf{r}, s_p) = \sqrt{\gamma_{j, s_e,
  s_p}(\textbf{r})} \, \psi^-_{j, s_e}(\textbf{r}) \psi^+_{s_p}(\textbf{r}).
\end{equation}
Note that, in general, it is justified to consider the
positron wavefunction to be collinear even though the electronic states are not.
Indeed, electron-positron spin-spin interactions are small and generally
neglected in PAS studies and positrons stay too far away from the nuclei to
experience any significant spin-orbit interaction. We thus assume that the
orbital part of the positron wavefunction is independent of the chosen
spin-polarization: $\psi^{+}_{s_p}(\textbf{r})~= \psi^{+}(\textbf{r})
\chi_{s_p}$, where $\chi_{s_p}$ denotes a two-component spinor for the positron.
Note that for the calculation of the momentum density from
eqs.~\eqref{eq:md_total} and \eqref{eq:md_magnetization}, we have to set
$\psi^+_{\uparrow}(\textbf{r}) = \psi^+_{\downarrow}(\textbf{r})$ instead of
explicitly setting a polarization. \\
In our calculations, we consider the state-dependent enhancement
factors~\cite{Alatalo1996, *Barbiellini1997}: $\gamma_{j, s_e, s_p} =
\lambda_{j, s_e, s_p}^{LDA} / \lambda_{j, s_e, s_p}^{IPM}$. The $\lambda$'s
denote the partial annihilation
rates in the LDA and independent particle model (IPM), respectively, and the
former is calculated as:
\begin{equation} \label{eq:PartialAnnihilationRate}
  \lambda_{j, s_e, s_p}^{LDA} = \pi r_e^2 c \int d\textbf{r} \, | \psi_{j,
    s_e}^-(\textbf{r}) |^2 | \psi_{s_p}^+(\textbf{r}) |^2
    \gamma(n^-(\textbf{r})),
\end{equation}
with $\gamma(n^-(\textbf{r}))$ the LDA enhancement factor parametrized by
Drummond~\cite{Drummond2011}. The IPM annihilation rates are obtained by setting
$\gamma(n^-(\textbf{r})) = 1$.\\
The high-momentum components of the wavefunctions are important to accurately
calculate the electron-positron momentum density. It is thus necessary to use
the all-electron wavefunctions in the above formulae instead of the soft pseudo
wavefunctions, i.e.\ we explicitly perform the PAW
transformation~\cite{Blochl1994}:
\begin{equation}
\vert \psi^- \rangle = \vert \tilde{\psi}^- \rangle + 
  \sum_i \left( \vert \phi^-_i \rangle - \vert \tilde{\phi}^-_i \rangle \right)
  \langle \tilde{p}_i \vert \tilde{\psi}^- \rangle.
\end{equation}
Here, $\vert \tilde{\psi}^-_j \rangle$ are the soft pseudowavefunctions,
$\langle \tilde{p}_i \vert$ are the projectors and $\vert \phi_i^- \rangle$ and
$\vert \tilde{\phi}_i^- \rangle$ are the localized all-electron and soft pseudo
partial waves of the ions respectively.  The details on how we performed this
transformation can be found in refs.~\onlinecite{Makkonen2005, Makkonen2006}.\\

\subsection{Positronium model}
\begin{figure}
\includegraphics[width=86mm]{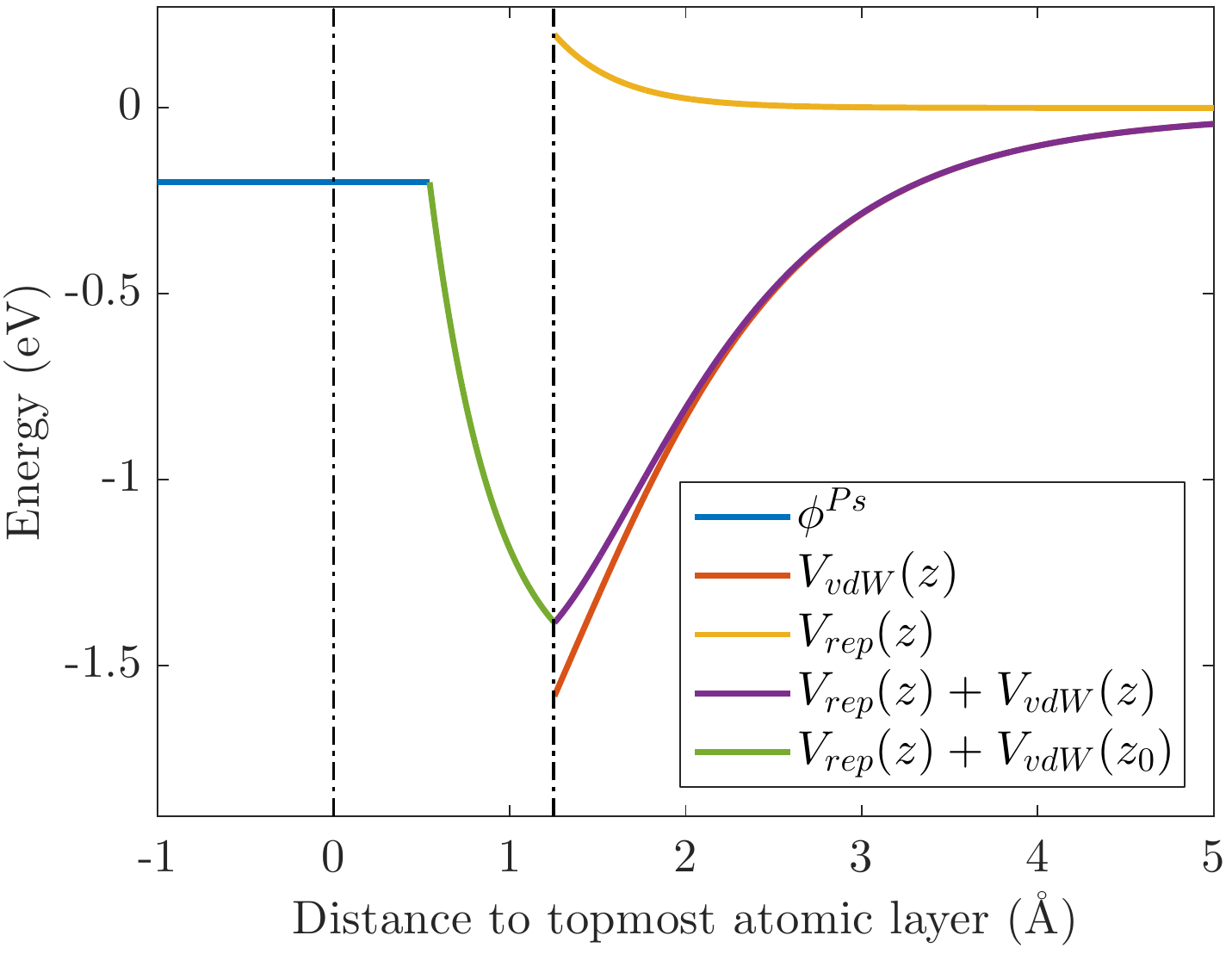}
\caption{\label{fig:PsPotential}  Potential obtained for the Ps
model with the values for $C$, $z_0$, $\lambda$ and $\phi^+$ mentioned in the
discussion. For the electronic workfunction, we took $\phi^- = 4.612~\text{eV}$,
which gives a Ps workfunction of $\phi^{Ps} = 0.2~\text{eV}$ and lies in the
middle of the range of values for which the model gives an activation energy in
good agreement with the experimental result.}
\end{figure}

We can theoretically determine the activation energy for Ps desorption from a
Bi$_2$Te$_2$Se, of which the experimental results are described in
ref.~\onlinecite{Shastry2015}, by calculating the particle's binding energy to
the surface. In order to model the Ps state, we consider the
Schr\"odinger equation for a neutral particle in an effective potential
well~\cite{Saniz2007}.  Here, the effective potential outside the surface is
determined by an attractive and a repulsive contribution. The repulsive
contribution, due to the overlap of the electron of the Ps with electrons of the
material, is given by
\begin{equation}
V_R(z) = \vert \phi^{Ps} \vert e^{-(z - z_0)/ \lambda},
\end{equation}
where $\phi^{Ps}$ is the Ps workfunction, $z_0$ the background edge position and
$\lambda$ the characteristic length of the electron density decay outside the
surface. The Ps workfunction can be calculated by taking the sum of the
workfunctions of the constituent particles minus their binding energy:
$\phi^{Ps} = \phi^+ + \phi^- - 0.25 \, \text{Ha}$. The attractive part of the
interaction is given by the Van der Waals interaction and can be written as
\begin{equation}
V_{vdW}(z) = - \frac{C}{(z - z_0')^3} F( (z-z_0')/\lambda),
\end{equation}
where the strength of the interaction is given by the
expression~\cite{Zaremba1976}:
\begin{equation}
C = \frac{\hbar}{4 \pi} \int \limits_0^\infty d \omega \, \alpha(i \omega) 
  \left( \frac{ \epsilon(i \omega) - 1}{\epsilon(i \omega) + 1} \right).
\end{equation}
The bulk dielectric function $\epsilon$ at imaginary frequencies can be obtained
by first evaluating the dielectric function at real frequencies, which is
readily calculated from first-principles in the RPA approximation, and then
applying analytic continuation. The Ps polarizability $\alpha$ can be obtained
from the analytic expression for H-like atoms, given in
ref.~\onlinecite{Szmytkowski2001}, by rescaling.  Indeed, the Ps problem can be
solved by going to the center of mass coordinates, which then yield the same
equations as for the H atom. The only differences are that the Bohr radius is
twice as large and the ionization energy is half the value of that of H. The
analytic damping function $F$, for which we take expression (17) of
ref.~\onlinecite{Patil2002}, describes the saturation of the Van der Waals
interaction as the particle draws closer to the surface and regularizes the
divergence at the reference plane position $z=z_0'$. The reference plane
position can in principle take another value than the background edge position
but since they are both, in the case of an elementary metal with lattice
parameter $a$, located close to $a/2$, we make the approximation $z_0' = z_0$.
For $z < z_0$, we extend the repulsive interaction, and add $V_{vdW}(z_0)$ to
ensure the continuity of the potential, with a cutoff set by the Ps
workfunction:
\begin{equation} \begin{split}
V(z) = \, & \min \{ \phi^{Ps}, V_R(z) + V_{vdW}(z_0) \} \Theta(z < z_0) \\
  & + \{ V_R(z) + V_{vdW}(z) \} \Theta(z \ge z_0).
\end{split} \end{equation}
The different contributions to the potential are show in
figure~\ref{fig:PsPotential}.  The Ps state and its energy are obtained by
solving the resulting Schr\"odinger equation
\begin{equation}
-\frac{\psi''}{4} + V(z) \psi = E \psi.
\end{equation}

\section{\label{section:ResultsTheory} Computational results}
\begin{figure} 
\includegraphics[width=86mm]{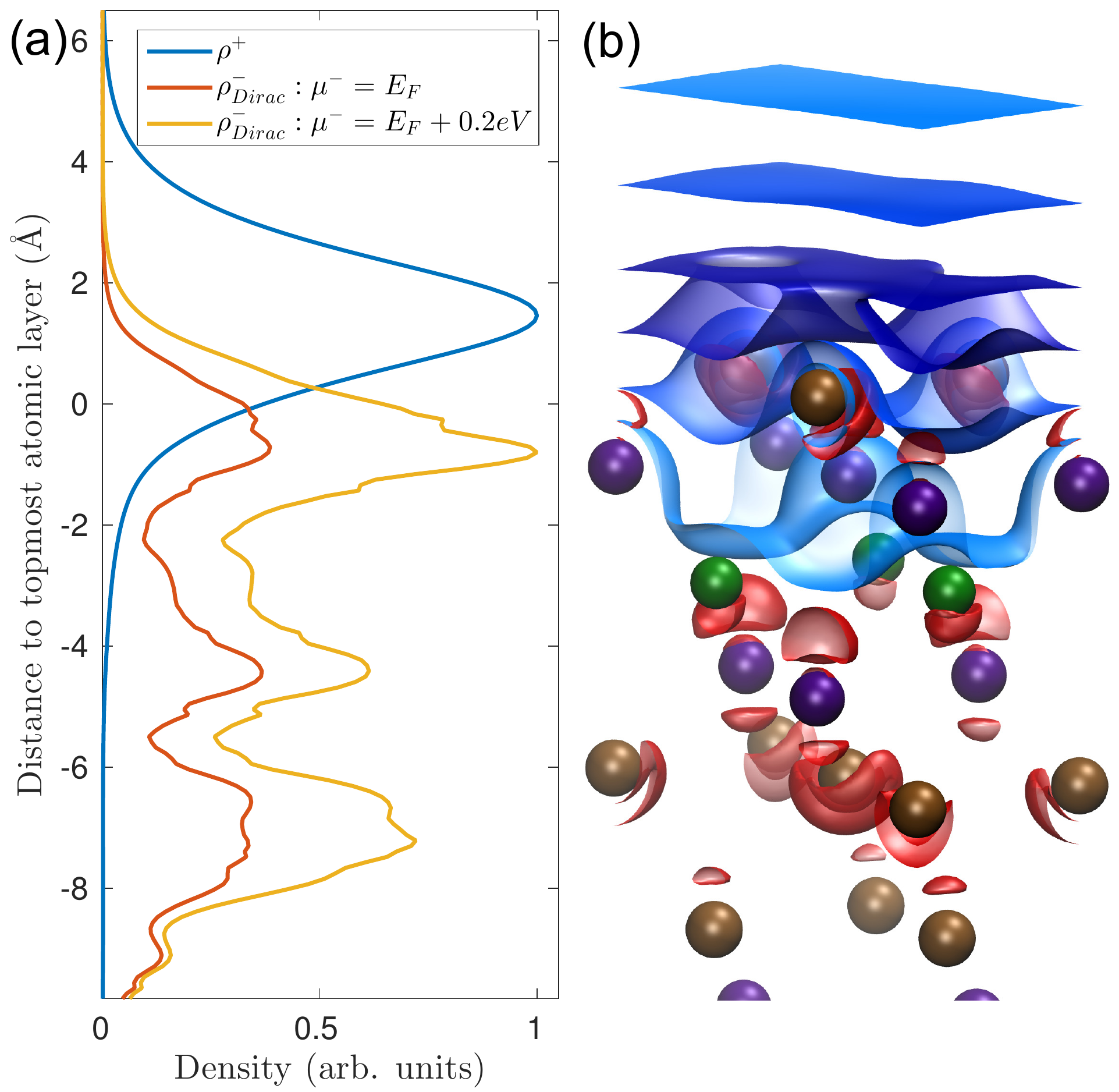}
\caption{\label{fig:Overlap}  Overlap of the positron SS with the
Dirac states.  (a) Planar average of the positron (blue) and electron
(red/yellow) density associated with the Dirac states below the Fermi-energy for
two different values of the chemical potential $\mu^-$. (b) Density of the
topological surface state and the positron in the same spatial region as panel
(a). The progressively lighter blue isosurfaces show the positron density at
80\%, 20\% and 2\% of the maximum value, respectively, and the red isosurfaces
show the electronic charge density associated with the electron states on the
Dirac cone below the Fermi-level at 10\% of the maximum value. The Bi, Te and Se
atoms are shown in purple, brown and green colors, respectively.} 
\end{figure}

\begin{figure*} 
\includegraphics[width=170mm]{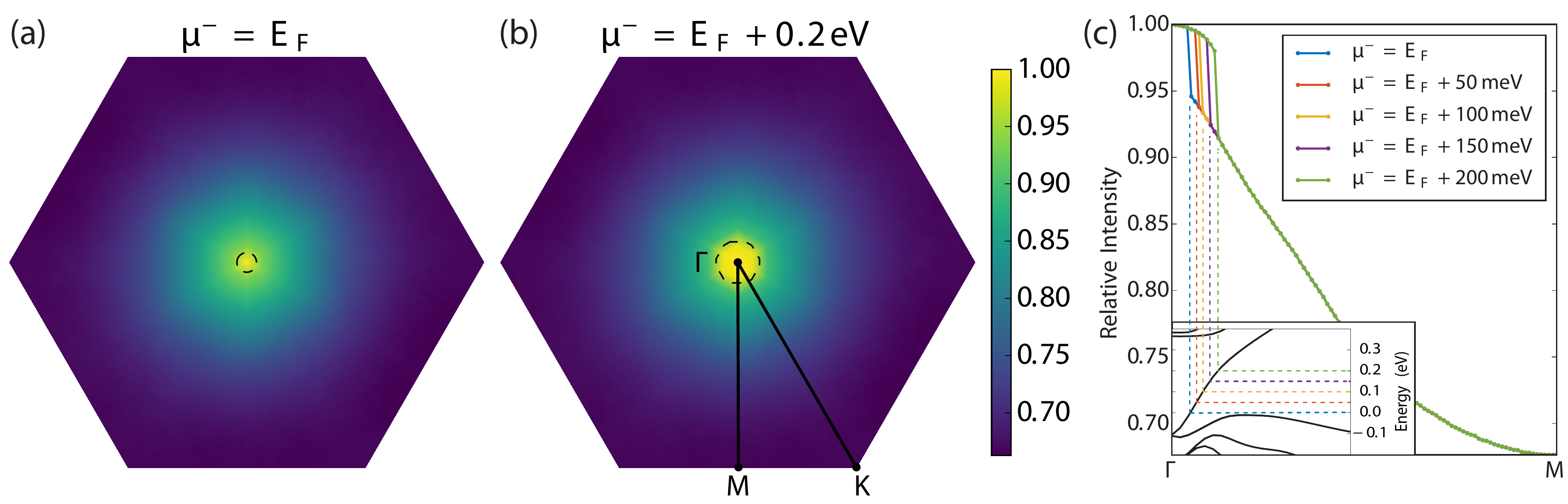}
\caption{\label{fig:LCWmap}  Theoretical momentum densities. (a)
LCW map with the chemical potential located at the Fermi-level. (b) LCW map with
the chemical potential raised by $0.2 \, \text{eV}$. The dashed line denote the
location of the Fermi-surface as derived from the electronic bandstructure. (c)
High resolution cuts through the LCW map along the $\Gamma$-$M$ direction for
different values of the chemical potential. The inset shows the band structure
near the Fermi-level $(E_F = 0.0 \, \text{eV})$.} 
\end{figure*}

\begin{figure}
\includegraphics[width=86mm]{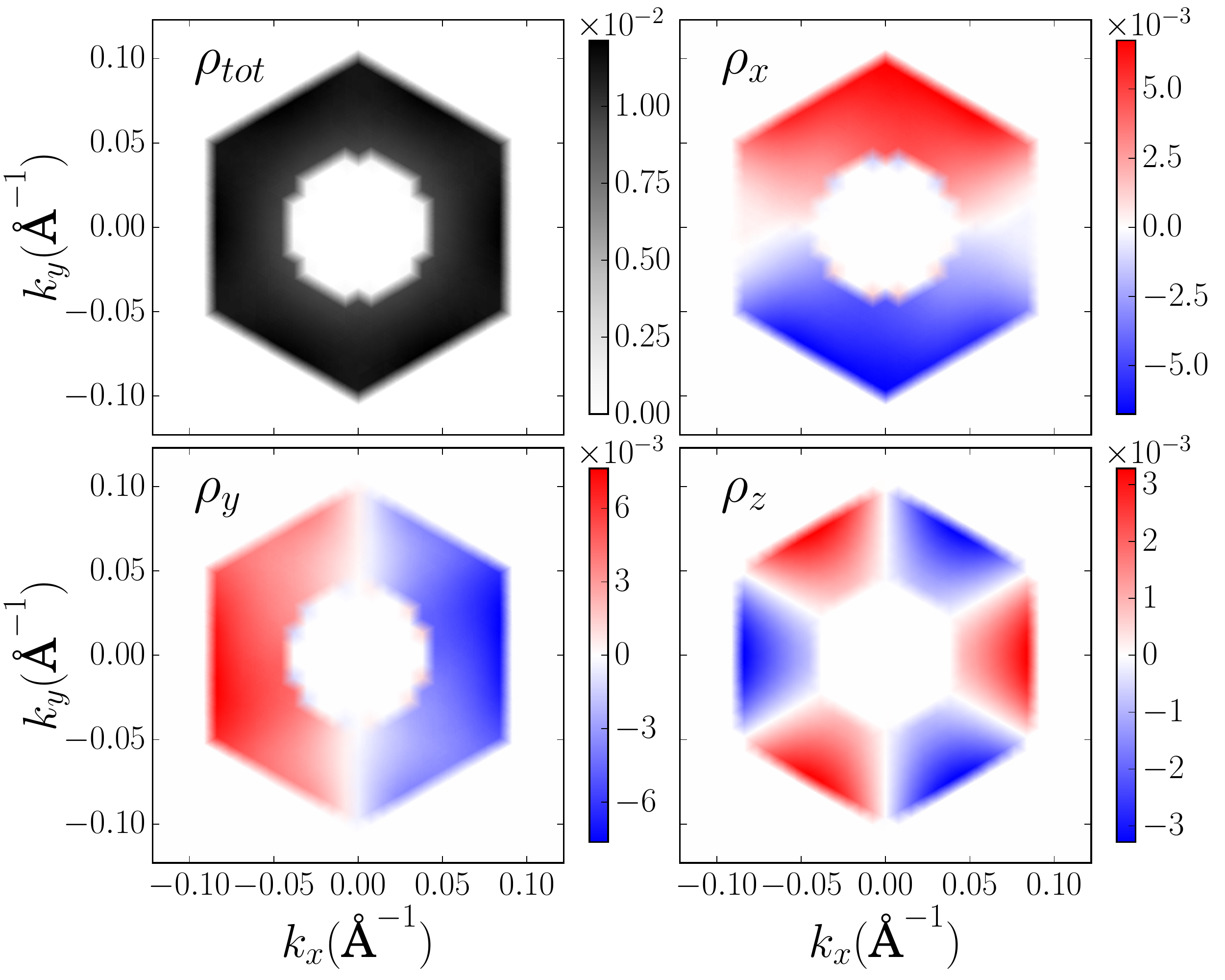}
\caption{\label{fig:MagneticLCW}  Difference between the LCW maps
obtained with different doping levels of the Dirac cone: $\mu^- = E_F + 0.2 \,
\text{eV}$ and $\mu^- = E_F$. The top left pane of the figure show the total
amplitude of the LCW map. The top right, bottom left and bottom right figures
show the magnetization components along the $x$-, $y$- and $z$-axes
respectively. We only show the result zoomed in around the $\Gamma$-point as the
difference between the LCW maps is exactly zero in the rest of the Brillouin
zone. The inner- and othermost edges of the non-zero part in the plots
correspond with the dashed lines shown in figures \ref{fig:LCWmap}(a) and
\ref{fig:LCWmap}(b), respectively. The length of the reciprocal axes is $|
\textbf{b}| = 1.688 \, \text{\AA}^{-1}$ and the amplitudes are given in
$\text{ps}^{-1} \text{\AA}^{2}$. (It is readidly seen that the units of
the LCW map are in $\text{ps}^{-1} \text{\AA}^2$ by realizing that the integral
over the LCW map yields the positron's annihilation rate, or, in the case of the
magnetic LCW maps, the difference in annihilation rate between two measurements
with opposite spin polarizations for the positron.)}
\end{figure}

We start our discussion of the computations by showing that the measured Ps
activation energy $E_a = 0.4~\text{eV}$~\cite{Shastry2015} is consistent with
the theoretical predictions. We take the activation energy to be equal to the
groundstate energy predicted by the Ps model discussed in the previous section.
For the parameters in the model, we find that the Van der Waals interaction
strength evaluates to $C = 2.306~\text{eV} \cdot \text{\AA}^3$ and from the
electronic and positronic workfunctions $\phi^- = 4.904~\text{eV}$ and $\phi^+ =
2.392~\text{eV}$, we obtain $\phi^{Ps} = 0.493~\text{eV}$. The values for the
background edge position and the characteristic length of the electron density
decay in the vacuum region are given by $z_0 = 1.250~\text{\AA}$ and $\lambda =
0.365~\text{\AA}$. Using these values, the model predicts that the Ps forms a
delocalized state in the bulk of the material. We note, though, that the
experimental value for the electronic workfunction $\phi^- = 4.5~\text{eV}$ is
lower than the theoretical one. It is thus sensible to consider the outcome of
the model for $\phi^- \in [4.5, 4.9]~\text{eV}$. Over the range $\phi^- =
4.90~\text{eV}$ to $\phi^- = 4.72~\text{eV}$, we find that the groundstate
gradually shifts from the bulk to the surface. To determine when we have a
surface state, we set the criterion that the Ps density should decay below 1\%
of its maximum value beyond the first QL block inside the material. In the range
$\phi^- \in [4.52, 4.72]~\text{eV}$, the Ps model predicts a surface state with
a binding energy of $E_{Ps} = 0.40 \pm 0.05~\text{eV}$, in good agreement with
the experimental results.\\ 
Next, we investigate the predictions of the 2CDFT calculations to determine
whether they support the proposed interpretation of the PAES and AMPS
experiments.  Our first observation is that the positron in its groundstate
indeed resides in the surface's image potential well rather than the gaps in
between the QLs, which also act as strong positron traps. We obtain the binding
energy of the positron by taking the difference between the vacuum level and the
positron's chemical potential. The vacuum level is determined in the usual
way by the taking the value of the Hartree potential in middle of the vacuum
region. We find that the positron SS has a binding energy of $E_b =
2.69~\text{eV}$, in excellent agreement with the measured value. We find that
the lifetime evaluates to $\tau = 309.25 \, \text{ps}$. This value
seems reasonable compared with the lifetime of $340 - 380 \, \text{ps}$ measured
for positrons trapped at the surface of colloidal PbSe quantum
dots~\cite{Chai2013}. On the other hand, a lifetime of $580 \, \text{ps}$ has
been determined for positrons trapped at an Al surface~\cite{Lynn1984}, which
can not be reproduced within the LDA approximation~\cite{Nieminen1983}. One
workaround suggested in literature is to set the enhancement factor to zero for
$z > z_0$, i.e.\ assume that the positron will not annihilate in the vacuum
region~\cite{Nieminen1984}. We find, though, that this operation makes the
result for the lifetime depend sensitively on the value for the image potential
reference plane $z_0$. For this reason, as well as the scarcity of experimental
data that show this operation is justified, the rest of our calculations have
been carried out without modifying the LDA enhancement factor. \\
Now that the calculations confirmed the existence of the bound positron SS, we
turn to the important question of the extent to which this SS overlaps with the
Dirac cone electrons. This overlap is of central importance because it
determines the annihilation rate of the positron with the electrons occupying
the topological states and thus the sensitivity with which positron annihilation
spectroscopy can probe the Dirac states. This can be seen from eq.\
\eqref{eq:PartialAnnihilationRate}, where the partial annihilation rate is
determined by the sum over $\lambda_j$ where $j$ denotes the states on the cone.
\\
The computed densities of the positron SS, $\rho^+$, and the topological Dirac
states, $\rho^-_\text{Dirac}$, are shown in figure~\ref{fig:Overlap}. The
density of the topological states is obtained by summing the one-particle
densities for all states on the cone between the Dirac point and a specific
value for the electron chemical potential $\mu^-$. Although the positron is seen
to probe only the topmost atomic layers of the material, it still penetrates the
material sufficiently to have a significant overlap with the Dirac states.
Moreover, the left panel of figure~\ref{fig:Overlap} shows that the overlap with
the Dirac states changes sensitively depending on the population of the Dirac
states near the Fermi-level. Our calculations of the momentum density, discussed
below, further demonstrate that this underlying overlap translates into a clear
signal coming from the annihilation of the positron with the Dirac fermions. \\
A partially filled energy band when it crosses the Fermi energy gives rise to a
break in the electron momentum density, which is the basis of the measurement of
Fermi surfaces in materials via 2D-ACAR experiments. A standard procedure for
enhancing the Fermi surface signal in the spectrum is the Lock-Crisp-West (LCW)
map obtained by folding all the higher momentum (Umklapp) contributions into the
first Brillouin zone~\cite{Lock1973}. Figure~\ref{fig:LCWmap} shows the
calculated LCW map together with a cut along $\Gamma - M$ over a range
of values of the electron chemical potential, which simulates different doping
levels of the Dirac cone. The evolution of the plateau around the $\Gamma$-point
clearly indicates the sensitivity of the positron to the Dirac cone states. The
relative drop in intensity between $5\% - 7\%$ at the Fermi momentum compares
favorably with, for example, the 1\% drop found for the
Nd$_{2-x}$Ce$_{x}$CuO$_{4-\delta}$ high $T_c$ superconductor in which 2D-ACAR
experiments have been shown previously to be viable in detecting Fermi surface
sheets due to Cu-O planes~\cite{Barbiellini1995, *Shukla1996}.\\
A topic which has drawn considerable interest in the case of topological
insulators, is the spin-momentum locking of the topological states. Measurements
using spin-polarised positron beams exploit the fact that a two photon decay is
only possible between electrons and positrons with opposite
spins~\cite{Berko1964}. In recent work, spin-effects in the electronic structure
of simple ferromagnets were observed using differences between the doppler
broadening of the annihilation radiation (DBAR) measured with positron aligned
parallel and anti-parallel to a polarizing magnetic field.~\cite{Kawasuso2012}.
In a similar ACAR experiment, Weber et al.~\cite{Weber2015} successfully
resolved the spin-dependent Fermi surface of the ferromagnetic Heusler compound
Cu$_2$MnAl.  This motivates us to investigate whether spin-polarised positrons
can be used to detect the spin-structure of the topological states at the
surface. The signal from the Fermi-surface can be extracted from the LCW map by
taking the difference between the signal obtained at different doping levels. In
figure~\ref{fig:MagneticLCW}, we show the results obtained by taking the
difference between the LCW maps obtained with $\mu^- = E_F + 0.2 \, \text{eV}$
and $\mu^- = E_F$ in the vicinity of the $\Gamma$-point.  As expected, we see
the plateau due to the extra occupation of the cone in the total amplitude. Our
results for the magnetization along the $x$- and $y$-directions, agree well with
the results obtained in several studies of various tetradymite
TIs~\cite{Henk2012,Basak2011,Lin2011,Wang2011}, which all predict a clockwise
rotation of the spin. We see that the $z$-component of the magnetization
increases gradually away from the $\Gamma$-point. This out of plane component
develops due to the hexagonal warping of the Dirac cone, as pointed out by
Fu~\cite{Fu2009}. We note that the difference in amplitude for the magnetic
components is quite pronounced w.r.t.\ to the Fermi-surface signal.  Indeed, we
find that the signal from the magnetization about half that of the Fermi-surface
signal obtainable with an unpolarized beam. This means that the magnetization
signal still constitutes a promising $2\% - 4\%$ of the total signal. We note,
though, that in real experiments, positron beams are not perfectly polarized, as
we have assumed in our calculations. Thus, in experiment, a proper weighting has
to be performed which will lead to a smaller signal. 

\section{\label{section:Conclusion} Conclusion and Outlook}
Our study establishes the existence of a positron surface state near the
topological insulator Bi$_2$Te$_2$Se. The results of our calculations show that
this surface state can be exploited as a spectroscopic characterization tool
for probing surfaces of topological materials. Since a significant fraction of
positrons annihilate with electrons occupying Dirac cone states, 2D-ACAR
experiments should be able to measure their momentum distribution with high
precision~\cite{Dugdale2013}, and thus obtain information concerning the nature
of the Dirac states which is complementary to that accessed through
angle-resolved photoemission, scanning tunnelling and other surface-sensitive
spectroscopies without complications of related matrix element
effects~\cite{Bansil1999, *Campuzano1991, *Nieminen2009}. PAES and Doppler
broadening of the annihilation radiation~\cite{Tuomisto2013} measurements can,
in turn, be used to characterize the chemical composition of surfaces. In
combination with 2D-ACAR experiments, these positron spectroscopies could be
exploited to determine effects of various surface impurities on the topological
states, in addition to the role of bulk defects~\cite{Devidas2014}. Now our
study identified a positron surface state, positron spectroscopies can prove
valuable for the characterization of nano-structured topological insulators.
Indeed, positrons have shown to act as effective self-seeking probes for
nano-crystal surfaces without requiring the preparation of single crystal
specimens~\cite{Eijt2006}, whereas the applicability of conventional
spectroscopic techniques is limited. Finally, our calculations show that the
spin-textures of the Dirac states should be accessible through 2D-ACAR
measurements using a spin-polarized positron beam since positrons predominantly
annihilate with electrons of the opposite spin~\cite{Berko1964, Kawasuso2012,
Weber2015}.\\

\section{Acknowledgements} 
I.~M.\ acknowledges discussions with M.~Ervasti and A.~Harju.\\
V.~C.\ and R.~S.\ were supported by the FWO-Vlaanderen through Project No.\ G.\
0224.14N. The computational resources and services used in this work were in
part provided by the VSC (Flemish Supercomputer Center) and the HPC
infrastructure of the University of Antwerp (CalcUA), both funded by the
Hercules Foundation and the Flemish Government (EWI Department). I.~M.\
acknowledges financial support from the Academy of Finland (projects 285809 and
293932). The work at Northeastern University was supported by the US Department
of Energy (DOE), Office of Science, Basic Energy Sciences grant number
DE-FG02-07ER46352, and benefited from Northeastern University's Advanced
Scientific Computation Center (ASCC) and the NERSC supercomputing center through
DOE grant number DE-AC02-05CH11231. K.~S.\ and A.~W.\ acknowledge financial
support from the National Science Foundation through grants DMR-MRI-1338130 and
DMR-1508719. D.~H.\ received financial support of the National Science
Foundation (grant ECCS-1402738). J.~S.~M.\ was supported by the STC Center for
Integrated Quantum Materials under NSF grant DMR-1231319, NSF DMR-1207469 and
ONR N00014-13-1-0301.  B.~A.~A. also acknowledges support from the LabEx
ENS-ICFP: ANR-10-LABX-0010/ANR-10-IDEX-0001-02 PSL.

\section{References}
\bibliography{references_Bi2Te2Se}

\clearpage
\onecolumngrid
\appendix
\section{Detailed derivation spin-resolved momentum density}

In this appendix, we give a detailed derivation of the momentum
density formulas given in the main manuscript. We start from the general
expression given in ref.~\onlinecite{Ryzhikh1999}, which defines the rate to
start with the ground-state with $N$ electron and a single positron $\Psi$ and
end up in the final state $\Phi$, with $N-1$ electrons and two photons with
total momentum $\bo{p}$:
\begin{equation} \begin{split}
\Lambda_\nu(\bo{p}) & = 4 \pi r_e^2 c \displaystyle \sum_{i=1}^N \sum_{s_i, s_p} 
  \bigg \vert
    \int d \tau \int d \bo{r}_i \int d \bo{r}_p \, 
      \Phi^*_\nu(\bo{x}_1, \dots, \bo{x}_{i-1}, \bo{x}_{i+1}, \dots, \bo{x}_N) 
      e^{-i \bo{p} \cdot \bo{r}_p} \\
    & \quad \times \hat{S}^s_i \,  \delta(\bo{r}_i - \bo{r}_p) \,
      \Psi(\bo{x}_1, \dots, \bo{x}_N; \bo{x}_p)
  \bigg \vert^2.
\end{split} \end{equation}
To keep notation in check, we define $\bo{x}_i = \{ \bo{r}_i, s_i \}$ to denote
the particle position and spin, and $d \tau = \prod_{j = 1, j \ne i}^{N}
d\bo{x}_j$ to represent integration over all the non-annihilating electron
coordinates, thus including a sum over the possible spin directions of the
particle. The contraint that only electron-positron pairs in a singlet state
contribute to the 2$\gamma$ annihilation is taken into account through the
operator $\hat{S}^s_i = 1 - \hat{S}^2/2$, with $\hat{S}_i^2$ the total spin
operator of electron $i$ and the positron. As we will show further on,
this operator projects the electron-positron pair on the singlet states of the
respective pairs. \\ 
Thanks to the anti-symmetry of the wavefunction, we can swap the electron
indices around such that the annihilating electron always has the label $N$. We
make use of the delta function to get perform the integration over the positron
coordinate to get:
\begin{equation}
\Lambda_\nu(\bo{p}) = 4 \pi r_e^2 c \, N \displaystyle \sum_{s_N, s_p} 
  \left \vert
    \int d \tau \int d \bo{r}_N \, 
      \Phi^*_\nu(\bo{x}_1, \dots, \bo{x}_{N-1})  \,
      e^{-i \bo{p} \cdot \bo{r}_N} \, 
      \hat{S}^s_N \,
      \Psi(\bo{r}_1, s_1, \dots, \bo{r}_N, s_N; \bo{r}_N, s_p)
  \right \vert^2.
\end{equation}
If we are not concerned with the precise final state $\Phi^*_\nu$ we end up
with, we can define the 2$\gamma$ transition rate $\rho(\bo{p}) = \sum_\nu
\Lambda_\nu(\bo{p})$ from $\Psi$ to some final state with a pair of photons of
total momentum $\bo{p}$. The completeness relation: 
\begin{equation}
\displaystyle \sum_\nu \Phi^*_\nu(\bo{x}_1, \dots, \bo{x}_{N-1})
  \Phi_\nu(\bo{x}'_1, \dots, \bo{x}'_{N-1}) = 
  \delta(\bo{x}_1 - \bo{x}'_1) \dots \delta(\bo{x}_{N-1} - \bo{x}'_{N-1}),
\end{equation}
then allows us to write:
\begin{equation}
  \rho(\bo{p}) = 4 \pi r_e^2 c N \displaystyle \sum_{s_e, s_p} 
  \int d \bo{x}_1 \dots \int d \bo{x}_{N-1}
    \left \vert
      \int d \bo{r}_N \, e^{i \bo{p} \cdot \bo{x}_N} \,
      \hat{S}^s_N \, \Psi(\bo{x}_1, \dots, \bo{r}_N, s_N; \bo{r}_N, s_p)
    \right \vert^2.
\end{equation}
We now recognize the electron-positron two-body reduced density matrix, defined
as~\cite{Makkonen2005, Zubiaga2016}:
\begin{equation} \begin{split}
\Gamma^{ep}(\bo{r}_e, s_e, \bo{r}_p, s_p; \bo{r}'_e, s_e', \bo{r}'_p, s_p') & = 
  N \int d\bo{x}_1 \dots \int d \bo{x}_{N-1} \,
    \Psi^*(\bo{x}_1, \dots, \bo{x}_{N-1}, \bo{r}_e, s_e; \bo{r}_p, s_p) \\
  & \quad \times  
    \Psi(\bo{x}_1, \dots, \bo{x}_{N-1}, \bo{r}'_e, s'_e; \bo{r}'_p, s'_p).
\end{split} \end{equation}
It is convenient to introduce the natural geminals, also called
pairing-wavefunctions, which diagonalize the above density matrix:
\begin{equation}
\Gamma^{ep}(\bo{r}_e, s_e, \bo{r}_p, s_p; \bo{r}'_e, s_e', \bo{r}'_p, s_p') = 
  \displaystyle \sum_j g_j \, \alpha^*_j(\bo{r}_e, s_e; \bo{r}_p, s_p)  \,
  \alpha_j(\bo{r}'_e, s_e'; \bo{r}'_p, s_p').
\end{equation}
We thus arrive at the general expression for the 2$\gamma$ momentum density:
\begin{equation} \label{eq:MDwSingletOp}
  \rho(\bo{p}) = 4 \pi r_e^2 c \displaystyle \sum_j g_j \sum_{s_e, s_p}
    \left \vert
      \int d \bo{r} \,  e^{-i \bo{p} \cdot \bo{r}} \, 
      \hat{S}^s \,  \alpha_j(\bo{r}, s_e; \bo{r}, s_p)
    \right \vert^2,
\end{equation}
where we dropped the now unnecessary label on the singlet operator. \\
Let us now examine the effect of the singlet projection operator. We have:
\begin{equation} \begin{split}
\hat{S}^s & = 1 - \frac{\hat{S}^2}{2}\\
  & = 1 - \frac{1}{2} \left( 
    \hat{S}^2_{s_e} + \hat{S}^2_{s_p} + 
    2 \, \hat{\bo{S}}_{s_e} \cdot \hat{\bo{S}}_{s_p} \right)  \\
  & = 1 - \frac{1}{2} \left( 
    \hat{S}^2_{s_e} + \hat{S}^2_{s_p} + 
    2 \, \left[ \hat{S}_{s_e,x} \hat{S}_{s_p,x} +
      \hat{S}_{s_e, y} \hat{S}_{s_p, y} + 
      \hat{S}_{s_e, z} \hat{S}_{s_p, z} \right] \right)
\end{split} \end{equation}
If we make use of:
\begin{equation} \begin{split}
  \hat{S}^2_{s_e} \ket{s_e, s_p} 
  & = \hat{S}^2_{s_p} \ket{s_e, s_p} 
    = \frac{3}{4} \ket{s_e, s_p} \\
  \hat{S}_{s_e, x} \hat{S}_{s_p, x} \ket{s_e, s_p}
  & = \vert s_e s_p \vert \ket{-s_e, -s_p} 
    = \frac{1}{4} \ket{-s_e, -s_p}  \\
  \hat{S}_{s_e, y} \hat{S}_{s_p, y} \ket{s_e, s_p}
  & = - s_e s_p \ket{-s_e, -s_p}  \\
  \hat{S}_{s_e, z} \hat{S}_{s_p, z} \ket{s_e, s_p}
  & = s_e s_p \ket{s_e, s_p},
\end{split}\end{equation}
then we find:
\begin{equation}
\hat{S}^s \ket{s_e, s_p} = \left( \frac{1}{4} - s_e s_p \right) 
  \left( \ket{s_e, s_p} - \ket{-s_e, -s_p} \right).
\end{equation}
It is obvious that this gives zero if the electron and positron spins have the
same value and the prefactor becomes $1/2$ when they are anti-parallel. Note
that this operator thus indeed projects on the singlet states $\pm (\ket{\updown} -
\ket{\downup})$.\\
After performing the sum over all electron and positron spins, we find for the
momentum density expression:
\begin{equation} \label{eq:MDSpinTerms} \begin{split}
  \rho(\bo{p}) & = 4 \pi r_e^2 c \displaystyle \sum_j g_j
    \left \vert
      \int d \bo{r} e^{-i \bo{p} \cdot \bo{r}}
       \frac{1}{\sqrt{2}} \left(\alpha_j(\bo{r}, \uparrow; \bo{r}, \downarrow) 
          - \alpha_j(\bo{r}, \downarrow; \bo{r}, \uparrow) \right) 
    \right \vert^2 \\
  & = 2 \pi r_e^2 c \sum_j g_j
    \int d\bo{r} \int d\bo{r}' e^{-i \bo{p} \cdot (\bo{r} - \bo{r}')}
      \Big[
        \alpha_j(\bo{r}, \uparrow; \bo{r}, \downarrow)
          \alpha^*_j(\bo{r}', \uparrow; \bo{r}', \downarrow) +
        \alpha_j(\bo{r}, \downarrow; \bo{r}, \uparrow)
          \alpha^*_j(\bo{r}', \downarrow; \bo{r}', \uparrow) \\
  & \quad -
        \alpha_j(\bo{r}, \downarrow; \bo{r}, \uparrow)
          \alpha^*_j(\bo{r}', \uparrow; \bo{r}', \downarrow) -
        \alpha_j(\bo{r}, \uparrow; \bo{r}, \downarrow)  
          \alpha^*_j(\bo{r}', \downarrow; \bo{r}', \uparrow) 
      \Big]. \\
\end{split} \end{equation}
In the rest of the derivation, it is more convenient to work with a spinor
representation for the geminals, which we define as:
\begin{equation}
  \alpha_j(\bo{r}_e; \bo{r}_p) =  
  \begin{pmatrix}
    \alpha_j(\bo{r}_e, \uparrow; \bo{r}_p, \uparrow) \\
    \alpha_j(\bo{r}_e, \uparrow; \bo{r}_p, \downarrow) \\
    \alpha_j(\bo{r}_e, \downarrow; \bo{r}_p, \uparrow) \\
    \alpha_j(\bo{r}_e, \downarrow; \bo{r}_p, \downarrow)
  \end{pmatrix}.
\end{equation}
Equation \eqref{eq:MDwSingletOp} then becomes:
\begin{equation} \label{eq:MDwSpinor} \begin{split}
\rho(\bo{p}) & = 4 \pi r_e^2 c \sum_j g_j
    \bigg \vert \int d\bo{r} \, e^{-i \bo{p} \cdot \bo{r}} \hat{S}^s
    \alpha_j(\bo{r}; \bo{r}) \bigg \vert^2 \\
  & = 4 \pi r_e^2 c \sum_j g_j
    \int d\bo{r} \int d\bo{r}' \, 
    e^{-i \bo{p} \cdot (\bo{r} - \bo{r}')}
    \left[ \hat{S}^s \alpha_j^{\dagger}(\bo{r}'; \bo{r}') \right]
    \left[ \hat{S}^s \alpha_j(\bo{r}'; \bo{r}') \right].
\end{split} \end{equation}
Indeed, making use of:
\begin{equation}
  \hat{S}^s \alpha_j(\bo{r}_e; \bo{r}_p) = \frac{1}{2}
  \begin{pmatrix}
    0 \\
    \alpha_j(\bo{r}_e, \uparrow; \bo{r}_p, \downarrow) - 
      \alpha_j(\bo{r}_e, \downarrow; \bo{r}_p, \uparrow) \\
    \alpha_j(\bo{r}_e, \downarrow; \bo{r}_p, \uparrow) -
      \alpha_j(\bo{r}_e, \uparrow; \bo{r}_p, \downarrow) \\
    0
  \end{pmatrix},
\end{equation}
it is straightforward to check that one obtains the same result asin equation 
\eqref{eq:MDSpinTerms}. \\
Next, if we assume that the geminals are collinear in the positron spin, we can
write:
\begin{equation}
  \alpha_j(\bo{r}_e; \bo{r}_p) = 
  \begin{pmatrix}
    \alpha_j(\bo{r}_e, \uparrow; \bo{r}_p) \\
    \alpha_j(\bo{r}_e, \downarrow; \bo{r}_p) \\
  \end{pmatrix}
  \otimes
  \chi_p,
\end{equation}
where $\otimes$ denotes a direct product, the remaining arrow indicates the
electron spin, and $\chi_p$ is the (position-independent) spinor for the
positron. For a positron fully polarized along the positive and negative
$z$-axis, respectively, we have the single particle spinors:
\begin{equation}
\chi_p^{z^+} = 
  \begin{pmatrix}
    1 \\ 0
  \end{pmatrix}
\quad
\text{and}
\quad
\chi_p^{z^-} = 
  \begin{pmatrix}
    0 \\ 1
  \end{pmatrix}.
\end{equation}
For the geminal spinor this gives:
\begin{equation}
  \alpha^{z^+}_j(\bo{r}_e; \bo{r}_p) =  
  \begin{pmatrix}
    \alpha_j(\bo{r}_e, \uparrow; \bo{r}_p, \uparrow) \\
    0 \\
    \alpha_j(\bo{r}_e, \downarrow; \bo{r}_p, \uparrow) \\
    0
  \end{pmatrix},
  \quad
  \text{and}
  \quad
  \alpha^{z^-}_j(\bo{r}_e; \bo{r}_p) =  
  \begin{pmatrix}
    0 \\
    \alpha_j(\bo{r}_e, \uparrow; \bo{r}_p, \downarrow) \\
    0 \\
    \alpha_j(\bo{r}_e, \downarrow; \bo{r}_p, \downarrow)
  \end{pmatrix}.
\end{equation}
and after applying the singlet operator to them:
\begin{equation}
  \hat{S}^s \alpha^{z^+}_j(\bo{r}_e; \bo{r}_p) = \frac{1}{2}
  \begin{pmatrix}
    0 \\
    0 \\
    \alpha_j(\bo{r}_e, \downarrow; \bo{r}_p, \uparrow) \\
    0
  \end{pmatrix},
  \quad
  \text{and} 
  \quad
  \hat{S}^s \alpha^{z^-}_j(\bo{r}_e; \bo{r}_p) = \frac{1}{2}
  \begin{pmatrix}
    0 \\
    \alpha_j(\bo{r}_e, \uparrow; \bo{r}_p, \downarrow) \\
    0 \\
    0 \\
  \end{pmatrix}.
\end{equation}
From \eqref{eq:MDwSpinor} we obtain:
\begin{equation} \begin{split}
  \rho_{z^+}(\bo{p}) & = 2 \pi r_e^2 c \sum_j g_j
    \int d\bo{r} \int d\bo{r}' \, e^{-i \bo{p} \cdot (\bo{r} - \bo{r}')}
    \alpha_j(\bo{r}, \downarrow; \bo{r}, \uparrow) 
    \alpha^*_j(\bo{r}', \downarrow;\bo{r}', \uparrow) \\
  \rho_{z^-}(\bo{p}) & = 2 \pi r_e^2 c \sum_j g_j
    \int d\bo{r} \int d\bo{r}' \, e^{-i \bo{p} \cdot (\bo{r} - \bo{r}')}
    \alpha_j(\bo{r}, \uparrow; \bo{r}, \downarrow)
    \alpha^*_j(\bo{r}', \uparrow;\bo{r}', \downarrow).
\end{split} \end{equation}
The magnetization is obtained as the difference between these two spectra
and gives:
\begin{equation}
\rho_z(\bo{p}) = \rho_{z^-}(\bo{p}) - \rho_{z^+}(\bo{p}) =
  2 \pi r_e^2 c \displaystyle \sum_j g_j 
  \left[
    \bigg \vert \int d\bo{r} e^{-i \bo{p} \cdot \bo{r}} 
      \alpha_j(\bo{r}, \uparrow; \bo{r}, \downarrow) \bigg \vert^2
  - \bigg \vert \int d\bo{r} e^{-i \bo{p} \cdot \bo{r}} 
      \alpha_j(\bo{r}, \downarrow; \bo{r}, \uparrow) \bigg \vert^2
  \right]
\end{equation}
A positron polarized along the $x$-axis is represented by the single particle 
spinors:
\begin{equation}
\chi_p^{x^+} = \frac{1}{\sqrt{2}}
  \begin{pmatrix}
    1 \\ 1
  \end{pmatrix}
\quad
\text{and}
\quad
\chi_p^{x^-} = \frac{1}{\sqrt{2}}
  \begin{pmatrix}
    1 \\ -1
  \end{pmatrix}.
\end{equation}
Thus we get:
\begin{equation}
\alpha^{x^+}_j(\bo{r}; \bo{r}) =  \frac{1}{\sqrt{2}}
\begin{pmatrix}
  \alpha_j(\bo{r}, \uparrow; \bo{r}, \uparrow) \\
  \alpha_j(\bo{r}, \uparrow; \bo{r}, \downarrow) \\
  \alpha_j(\bo{r}, \downarrow; \bo{r}, \uparrow) \\
  \alpha_j(\bo{r}, \downarrow; \bo{r}, \downarrow)
\end{pmatrix},
\quad
\alpha^{x^-}_j(\bo{r}; \bo{r}) = \frac{1}{\sqrt{2}}
\begin{pmatrix}
  \alpha_j(\bo{r}, \uparrow; \bo{r}, \uparrow) \\
  - \alpha_j(\bo{r}, \uparrow; \bo{r}, \downarrow) \\
  \alpha_j(\bo{r}, \downarrow; \bo{r}, \uparrow) \\
  - \alpha_j(\bo{r}, \downarrow; \bo{r}, \downarrow)
\end{pmatrix},
\end{equation}
and:
\begin{equation}
\hat{S}^s \alpha^{x^+}_j(\bo{r}; \bo{r}) =  \frac{1}{2 \sqrt{2}}
\begin{pmatrix}
  0 \\
  \alpha_j(\bo{r}, \uparrow; \bo{r}, \downarrow) - 
    \alpha_j(\bo{r}, \downarrow; \bo{r}, \uparrow) \\
  \alpha_j(\bo{r}, \downarrow; \bo{r}, \uparrow) -
    \alpha_j(\bo{r}, \uparrow; \bo{r}, \downarrow) \\
  0 \\
\end{pmatrix}, 
\quad
\hat{S}^s \alpha^{x^-}_j(\bo{r}; \bo{r}) = \frac{1}{2 \sqrt{2}}
\begin{pmatrix}
  0 \\
  - \alpha_j(\bo{r}, \uparrow; \bo{r}, \downarrow) -
    \alpha_j(\bo{r}, \downarrow; \bo{r}, \uparrow) \\
  \alpha_j(\bo{r}, \downarrow; \bo{r}, \uparrow) +
    \alpha_j(\bo{r}, \uparrow; \bo{r}, \downarrow) \\
  0 \\
\end{pmatrix},
\end{equation}
which result in the momentum densities:
\begin{equation} \begin{split}
  \rho_{x^+}(\bo{p}) & = 2 \pi r_e^2 c \sum_j g_j
    \int d\bo{r} \int d\bo{r}' e^{-i \bo{p} \cdot (\bo{r} - \bo{r}')}
      \frac{1}{2}\Big[
        \alpha_j(\bo{r}, \uparrow; \bo{r}, \downarrow)
          \alpha^*_j(\bo{r}', \uparrow; \bo{r}', \downarrow) +
        \alpha_j(\bo{r}, \downarrow; \bo{r}, \uparrow)
          \alpha^*_j(\bo{r}', \downarrow; \bo{r}', \uparrow) \\
  & \quad -
        \alpha_j(\bo{r}, \downarrow; \bo{r}, \uparrow)
          \alpha^*_j(\bo{r}', \uparrow; \bo{r}', \downarrow) -
        \alpha_j(\bo{r}, \uparrow; \bo{r}, \downarrow)  
          \alpha^*_j(\bo{r}', \downarrow; \bo{r}', \uparrow) 
      \Big]. \\
  \rho_{x^-}(\bo{p}) & = 2 \pi r_e^2 c \sum_j g_j
    \int d\bo{r} \int d\bo{r}' e^{-i \bo{p} \cdot (\bo{r} - \bo{r}')}
      \frac{1}{2}\Big[
        \alpha_j(\bo{r}, \uparrow; \bo{r}, \downarrow)
          \alpha^*_j(\bo{r}', \uparrow; \bo{r}', \downarrow) +
        \alpha_j(\bo{r}, \downarrow; \bo{r}, \uparrow) 
          \alpha^*_j(\bo{r}', \downarrow; \bo{r}', \uparrow) \\
  & \quad +
        \alpha_j(\bo{r}, \downarrow; \bo{r}, \uparrow)
          \alpha^*_j(\bo{r}', \uparrow; \bo{r}', \downarrow) +
        \alpha_j(\bo{r}, \uparrow; \bo{r}, \downarrow)  
          \alpha^*_j(\bo{r}', \downarrow; \bo{r}', \uparrow) 
      \Big]. \\
\end{split} \end{equation}
So the magnetization in this case is:
\begin{equation} \begin{split}
  \rho_{x}(\bo{p}) = \rho_{x^-}(\bo{p}) - \rho_{x^+}(\bo{p}) & =
  2 \pi r_e^2 c \sum_j g_j
    \int d\bo{r} \int d\bo{r}' e^{-i \bo{p} \cdot (\bo{r} - \bo{r}')} \\
    & \quad \times \Big[
     \alpha_j(\bo{r}, \downarrow; \bo{r}, \uparrow) 
        \alpha^*_j(\bo{r}', \uparrow; \bo{r}', \downarrow) +
      \alpha_j(\bo{r}, \uparrow; \bo{r}, \downarrow)  
        \alpha^*_j(\bo{r}', \downarrow; \bo{r}', \uparrow) 
    \Big]
\end{split} \end{equation}
Finally, a positron polarized along the $y$-axis is represented by the single
particle spinors:
\begin{equation}
\chi_p^{y^+} = \frac{1}{\sqrt{2}}
  \begin{pmatrix}
    1 \\ i
  \end{pmatrix}
\quad
\text{and}
\quad
\chi_p^{y^-} = \frac{1}{\sqrt{2}}
  \begin{pmatrix}
    1 \\ -i
  \end{pmatrix}.
\end{equation}
The geminal spinors become:
\begin{equation}
\alpha^{y^+}_j(\bo{r}; \bo{r}) =  \frac{1}{\sqrt{2}}
\begin{pmatrix}
  \alpha_j(\bo{r}, \uparrow; \bo{r}, \uparrow) \\
  i \alpha_j(\bo{r}, \uparrow; \bo{r}, \downarrow) \\
  \alpha_j(\bo{r}, \downarrow; \bo{r}, \uparrow) \\
  i \alpha_j(\bo{r}, \downarrow; \bo{r}, \downarrow)
\end{pmatrix},
\quad
\alpha^{y^-}_j(\bo{r}; \bo{r}) = \frac{1}{\sqrt{2}}
\begin{pmatrix}
  \alpha_j(\bo{r}, \uparrow; \bo{r}, \uparrow) \\
  -i \alpha_j(\bo{r}, \uparrow; \bo{r}, \downarrow) \\
  \alpha_j(\bo{r}, \downarrow; \bo{r}, \uparrow) \\
  -i \alpha_j(\bo{r}, \downarrow; \bo{r}, \downarrow)
\end{pmatrix},
\end{equation}
\begin{equation}
\hat{S}^s \alpha^{y^+}_j(\bo{r}; \bo{r}) =  \frac{1}{2 \sqrt{2}}
\begin{pmatrix}
  0 \\
  i \alpha_j(\bo{r}, \uparrow; \bo{r}, \downarrow) - 
    \alpha_j(\bo{r}, \downarrow; \bo{r}, \uparrow) \\
  \alpha_j(\bo{r}, \downarrow; \bo{r}, \uparrow) -
    i \alpha_j(\bo{r}, \uparrow; \bo{r}, \downarrow) \\
  0 \\
\end{pmatrix}, 
\quad
\hat{S}^s \alpha^{y^-}_j(\bo{r}; \bo{r}) = \frac{1}{2 \sqrt{2}}
\begin{pmatrix}
  0 \\
  - i \alpha_j(\bo{r}, \uparrow; \bo{r}, \downarrow) -
    \alpha_j(\bo{r}, \downarrow; \bo{r}, \uparrow) \\
  \alpha_j(\bo{r}, \downarrow; \bo{r}, \uparrow) +
    i \alpha_j(\bo{r}, \downarrow; \bo{r}, \uparrow) \\
  0 \\
\end{pmatrix},
\end{equation}
from which we find:
\begin{equation} \begin{split}
  \rho_{y^+}(\bo{p}) & = 2 \pi r_e^2 c \sum_j g_j
    \int d\bo{r} \int d\bo{r}' e^{-i \bo{p} \cdot (\bo{r} - \bo{r}')}
      \frac{1}{2}\Big[
        \alpha_j(\bo{r}, \uparrow; \bo{r}, \downarrow)
          \alpha^*_j(\bo{r}', \uparrow; \bo{r}', \downarrow) +
        \alpha_j(\bo{r}, \downarrow; \bo{r}, \uparrow) 
          \alpha^*_j(\bo{r}', \downarrow; \bo{r}', \uparrow) \\
  & \quad + 
        i  \alpha_j(\bo{r}, \downarrow; \bo{r}, \uparrow)
           \alpha^*_j(\bo{r}', \uparrow; \bo{r}', \downarrow) -
        i \alpha_j(\bo{r}, \uparrow; \bo{r}, \downarrow)  
          \alpha^*_j(\bo{r}', \downarrow; \bo{r}', \uparrow) 
      \Big]. \\
  \rho_{y^-}(\bo{p}) & = 2 \pi r_e^2 c \sum_j g_j
    \int d\bo{r} \int d\bo{r}' e^{-i \bo{p} \cdot (\bo{r} - \bo{r}')}
      \frac{1}{2}\Big[
        \alpha_j(\bo{r}, \uparrow; \bo{r}, \downarrow) 
          \alpha^*_j(\bo{r}', \uparrow; \bo{r}', \downarrow) +
        \alpha_j(\bo{r}, \downarrow; \bo{r}, \uparrow) 
          \alpha^*_j(\bo{r}', \downarrow; \bo{r}', \uparrow) \\ 
  & \quad -
        i \alpha_j(\bo{r}, \downarrow; \bo{r}, \uparrow) 
          \alpha^*_j(\bo{r}', \uparrow; \bo{r}', \downarrow) +
        i \alpha_j(\bo{r}, \uparrow; \bo{r}, \downarrow)  
          \alpha^*_j(\bo{r}', \downarrow; \bo{r}', \uparrow)
      \Big]. \\
\end{split} \end{equation}
This gives the final component of the magnetization:
\begin{equation} \begin{split}
  \rho_{y}(\bo{p}) = \rho_{y^-}(\bo{p}) - \rho_{y^+}(\bo{p}) & = 
   2 i \pi r_e^2 c \sum_j g_j
    \int d\bo{r} \int d\bo{r}' e^{-i \bo{p} \cdot (\bo{r} - \bo{r}')} \\
    & \quad \times \Big[
      \alpha_j(\bo{r}', \uparrow; \bo{r}', \downarrow) 
        \alpha^*_j(\bo{r}, \downarrow; \bo{r}, \uparrow) -
      \alpha_j(\bo{r}', \downarrow; \bo{r}', \uparrow) 
        \alpha^*_j(\bo{r}, \uparrow; \bo{r}, \downarrow) 
    \Big].
\end{split} \end{equation}\\
If we introduce the notation:
\begin{equation}
A_{j, s_e, s_p}(\bo{p}) = \int d \bo{r} \, e^{-i \bo{p} \cdot \bo{r}} 
  \alpha_j(\bo{r}, s_e; \bo{r}, s_p)
\end{equation}
and the matrix:
\begin{equation}
\Gamma_j(\bo{p}) = \begin{pmatrix}
  | A_{j, \updown}(\bo{p}) |^2 & 
    A_{j, \updown}(\bo{p}) A_{j, \downup}^*(\bo{p}) \\
  A_{j, \downup}(\bo{p}) A_{j, \updown}^*(\bo{p}) &
    | A_{j, \downup}(\bo{p}) |^2  
\end{pmatrix},
\end{equation}
then the above results can be written as:
\begin{equation}
\rho_i(\bo{p}) =  2 \pi r_e^2 c \displaystyle \sum_j g_j 
  \text{Tr} \left[ \sigma_i \Gamma_j(\bo{p}) \right],
\end{equation}
where $i = \{x, y, z\}$ and $\sigma_i$ are the Pauli matrices. \\
Let us now derive the momentum density as measured in experiments with
unpolarized positron beams. In this case, there is a \emph{statistical}
uncertainty on the direction of the positron spin. We can assume that the
50\% of the positrons are in the up state and 50\% in the down state, w.r.t.\
whatever direction of the quantization axis. This means we measure:
\begin{equation}
\rho_{\text{tot}}(\bo{p}) =
  \frac{1}{2} (\rho_{z^+}(\bo{p}) + \rho_{z^-}(\bo{p})) = 
  \pi r_e^2 c \displaystyle \sum_j g_j 
    \text{Tr} \left[ \Gamma_j(\bo{p}) \right],
\end{equation}
which results in the correct prefactor found in literature.\\

\end{document}